%% file: main.tex
\documentclass{article}

\usepackage{PRIMEarxiv}

\usepackage[utf8]{inputenc} 
\usepackage{kotex}
\usepackage[T1]{fontenc}    
\usepackage{hyperref}       
\usepackage{url}            
\usepackage{booktabs}       
\usepackage{amsfonts}       
\usepackage{nicefrac}       
\usepackage{microtype}      
\usepackage{lipsum}
\usepackage{fancyhdr}       
\usepackage{graphicx}       
\graphicspath{{media/}}     
\usepackage{listings}

\usepackage{amsmath}
\usepackage{stmaryrd}
\usepackage{tikz}
\usepackage{amssymb}
\usetikzlibrary{shapes.geometric, arrows.meta, positioning, calc, fit, backgrounds}
\usepackage{amsthm} 
\usepackage{tcolorbox}
\usepackage{xcolor}
\usepackage{listings}

\usepackage{algorithm}
\usepackage{algorithmic}

\lstdefinelanguage{json}{
    basicstyle=\normalfont\ttfamily,
    numbers=left,
    numberstyle=\scriptsize\color{gray},
    stepnumber=1,
    numbersep=8pt,
    showstringspaces=false,
    breaklines=true,
    frame=lines,
    backgroundcolor=\color{white},
    string=[s]{"}{"},
    comment=[l]{:\ },
    morecomment=[l]{,"}
}

\title{Neuro-Symbolic AI for Korean Criminal Law: Sentencing Prediction and Document Drafting
}

\author{
  Yeonseok Lee \\
  SLING AI Inc. \\
  Incheon, Republic of Korea \\
  \texttt{ylee@sling.ai.kr}
}

\begin{document}

\newcommand{\bigast}{\scalebox{1.5}{$\ast$}}
\newcommand{\bank}{\textsf{bank}}
\newcommand{\arr}{\textsf{arr}}
\newcommand{\cfu}{\textsf{cfu}}

\maketitle

\begin{abstract}
The Korean criminal justice system utilizes summary proceedings (\textit{guyaksik}) to expedite high-volume minor infractions, such as simple driving under the influence (DUI), unlicensed driving, and minor traffic casualties. Although this mechanism improves judicial throughput, processing these cases creates a substantial administrative burden for prosecutors, driving the need for automated systems that can precisely translate unstructured legal text into deterministic statutory outcomes. While recent Large Language Models (LLMs) excel at semantic extraction, their probabilistic nature inherently limits their reliability in Legal Judgment Prediction tasks. Specifically, when confronted with the arithmetic constraints of legal statutes, LLMs can produce hallucinations. Given that legal accountability permits virtually no tolerance for stochastic errors, purely neural architectures remain limited in their direct judicial applications. To address these limitations, we propose a Neuro-Symbolic framework that bridges unstructured legal facts with formal verification. Our architecture restricts the LLM exclusively to semantic extraction, while offloading statutory fine calculations to a Satisfiability Modulo Theories solver. This division of labor reduces hallucination risks during computation. Furthermore, we incorporate a Human-in-the-Loop verification scheme to preserve professional legal oversight. We formalize the 2026 Sentencing Guidelines for Traffic Offenses within this pipeline, demonstrating a deterministic approach to supporting summary indictments.
\end{abstract}

\input{sections/01_introduction}

\input{sections/02_background}

\input{sections/03_ingestion}

\input{sections/04_extraction}

\input{sections/05_solver}

\input{sections/06_generation}

\input{sections/07_conclusion}

\newpage
\bibliographystyle{unsrt}  
\bibliography{references}

\input{sections/08_appendix}

\end{document}

%% file: sections/01_introduction.tex

\section{Introduction}
\label{sec:intro}

The Korean criminal justice system relies on summary proceedings (\textit{guyaksik} / 구약식) to accelerate the adjudication of high-volume traffic offenses, specifically simple driving under the influence (DUI), driving without a license, and traffic accidents resulting in bodily injury. 
Although this procedural mechanism is designed to streamline administrative operations, legal practitioners—including prosecutors and judges—face overwhelming workloads that can degrade overall efficiency and case processing velocity. 
Consequently, there is a need for automated legal frameworks capable of precisely transforming unstructured prosecutorial documents into mathematically sound legal judgments.

Recent advances in Large Language Models (LLMs) have demonstrated exceptional capabilities in extracting information from dense legal texts. 
However, deploying pure LLMs for Legal Judgment Prediction (LJP) presents fundamental limitations due to their probabilistic, next-token generative nature. 
Calculating exact statutory fines requires strict adherence to mathematical constraints and rule-based reasoning—demands under which LLMs frequently exhibit hallucinations. 
In criminal law, where legal accountability demands complete determinism, this susceptibility to hallucination renders isolated neural architectures insufficient.

To bridge the gap between unstructured natural language understanding and strict formal logic, we propose a Neuro-Symbolic framework that maps legal facts directly to formal mathematical specifications. 
By restricting the LLM to semantic extraction and delegating statutory calculations to a formal verification solver, our architecture establishes a pipeline for summary indictments. 

Specifically, the core contributions of this paper are threefold:
\begin{itemize}
    \item We design a neuro-symbolic pipeline that transforms unstructured police and prosecutorial records (e.g., breathalyzer logs and circumstantial reports) into deterministic legal outcomes, thereby reducing the risk of LLM hallucination during fine computation.
    \item We incorporate a Human-in-the-Loop (HITL) verification scheme that transparently anchors extracted legal variables to source reference texts, enhancing interpretability and maintaining human oversight.
    \item We construct a formal legal logic framework using the Z3 Satisfiability Modulo Theories (SMT) solver \cite{de2008z3} to compute precise sentencing ranges aligned with the 2026 Sentencing Guidelines for Traffic Offenses \cite{sc_korea_traffic_2026}.
\end{itemize}

%% file: sections/02_background.tex

\section{Background and Related Work}
\label{sec:background}

The proposed architecture relies on the intersection of South Korean criminal procedure, Legal Judgment Prediction (LJP), and formal verification systems. This section outlines the structural foundations and academic precedents that inform our methodology.

\subsection{Summary Proceedings in Traffic Crimes}

In South Korea, summary proceedings (\textit{guyaksik} / 구약식) provide an expedited legal track for minor offenses, such as traffic violations, unlicensed driving, and non-fatal accidents. 
This mechanism allows prosecutors to seek monetary fines via summary orders based on documentary evidence—such as police reports and breathalyzer results—without a formal trial. 
Given the high volume of traffic offenses, this process is vital for judicial efficiency. 
Automating information extraction and fine calculation in this domain could  reduce the administrative burden on legal practitioners, preventing minor cases from congesting the prosecutorial pipeline.

\subsection{Symbolic Approaches in Legal Reasoning}
Symbolic legal AI relies on explicit rule bases to represent statutory interpretation and procedural logic. PROLEG \cite{satoh2010proleg} implements the Japanese ``Presupposed Ultimate Fact'' theory in Prolog, separating core statutory rules from exceptions to model the legal burden of proof as adversarial dialogues. 
To make defeasible logic accessible beyond Prolog, PYTHEN \cite{thanh2026pythen} translates these concepts into a Python-native, JSON-based framework designed to interface directly with LLMs while seamlessly integrating with the broader Python data science ecosystem.

Addressing statutory revision, Fungwacharakorn and Satoh \cite{fungwacharakorn2022toward} apply Inductive Logic Programming (ILP) to resolve ``legal debugging'' problems, aligning legal canons of construction with top-down and bottom-up rule learning. 
Finally, empirical benchmarks by Robaldo et al.~\cite{robaldo2024compliance} demonstrate that while explainable systems like PROLEG provide transparent reasoning traces, Answer Set Programming (ASP) solvers scale far better on large datasets~\cite{robaldo2024compliance}. This trade-off between interpretability and computational performance highlights the challenge of integrating explainable formal reasoning with high-performance computational scalability in legal applications.

\subsection{LLMs in Korean Legal Text Processing and Analysis}
Recent work in Korean legal NLP has expanded Large Language Model (LLM) applications across information extraction, precedent analysis, and legal reasoning. 
Hwang et al. \cite{hwang2022data} showed that generative language models can efficiently extract structured case facts and sentencing details from unstructured precedents with minimal training data.
To establish domain-wide benchmarks, Hwang et al. \cite{hwang2022multi} introduced LBox Open, covering precedent corpus construction, text classification, summarization, and legal judgment prediction.
Moving toward improving legal accessibility for the general public in real-world consultation scenarios, Lee et al. \cite{lee2026kolegalqa} presented KoLegalQA, an expert-verified legal question-answering dataset designed to enforce statutory citation grounding and structured legal explanations.
Together, these works suggest that while LLMs show promising potential in understanding Korean legal texts, achieving fully grounded and verifiable legal AI remains an ongoing challenge.

\subsection{Neuro-Symbolic AI in Legal Reasoning}

To overcome the probabilistic limitations of pure LLMs, researchers are increasingly turning to Neuro-Symbolic AI, a paradigm that hybridizes the semantic pattern recognition of neural networks with the rigorous, rule-based guarantees of symbolic logic. 
A central component of this symbolic grounding relies on Satisfiability Modulo Theories (SMT) solvers, such as Z3 \cite{de2008z3}, which evaluate complex Boolean and mathematical constraints to mathematically prove or disprove logical satisfiability.

Recent scholarship has turned to neuro-symbolic methods to address the unpredictability and opacity of standard neural models in high-stakes areas like international tax law. 
From a jurisprudential standpoint, Hude \cite{hude2025toward} establishes a theoretical framework for legal neuro-symbolic AI rooted in Hartian legal positivism. 
This approach balances formal legal rules with interpretive flexibility by letting neural and symbolic techniques complement one another. 
Under this model, ``easy cases'' with straightforward statutory rules are processed through deterministic logic engines. 
Conversely, neural models handle ``hard cases,'' simulating a form of bounded judicial discretion to address interpretive gaps when explicit rules run out.

To implement this hybrid approach, recent studies employ multi-agent architectures that pair neural extraction with formal solvers. 
Sadowski and Chudziak \cite{sadowski2025verifiable} introduced SOLAR, a framework that decouples legal reasoning into two distinct phases: specialized LLM agents formalize statutory provisions into structured terminological ontologies (TBox), while a symbolic inference agent employs an SMT solver to apply rules to factual assertions (ABox) for deriving logically entailed conclusions, which are subsequently processed by a dedicated program interpreter to execute complex statutory calculations.

Similarly, Chen Linze and Cai Yufan \cite{chen2025towards} developed L4L, a framework that assigns adversarial roles to prosecutor and defense LLM agents for suspect-centric fact-tuple extraction. 
Instead of utilizing a deterministic adjudicator to filter out arguments, L4L leverages the Z3 SMT solver to rigorously verify statutory applicability and clause-level qualification; if an inconsistency is detected, an iterative feedback loop passes the minimal unsatisfiable core back to an autoformalizer to dynamically revise the legal constraints before a judge LLM renders the final verdict.

In the specific domain of Legal Judgment Prediction (LJP), coupling LLMs with Satisfiability (SAT) solvers significantly mitigates the risk of semantic hallucinations. Liu et al. \cite{liu2026enhancing} demonstrated that translating natural language case facts and statutory elements into formal logic queries for Z3 reasoning allows systems to ignore semantic distractor features and achieve high evaluation accuracy on constitutive legal elements. 

Zhang et al. \cite{zhang2025rljp} proposed RLJP, a framework that initializes First-Order Logic (FOL) judgment rules via a LLM agent using legal provisions and precedents, and dynamically refines them through confusion-aware contrastive learning based on tree-splitting operations. By combining lightweight semantic prescreening for candidate label generation with formal FOL rule execution, their method systematically addresses the rigidity of fixed legal logic to derive precise articles, charges, and prison terms, particularly showcasing superior performance in complex and lengthy case narratives.

Our work extends this field by proposing an application that bridges Korean criminal records and the 2026 Sentencing Guidelines \cite{sc_korea_traffic_2026}. 
By confining the LLM to a semantic extraction role and delegating the statutory calculation to a formal Z3 proof, our theoretical framework can establish a mathematically sound pipeline that neutralizes generative hallucination risks within prosecutorial automation workflows.

%% file: sections/03_ingestion.tex

\section{Raw Data Ingestion (Module 1)}
\label{sec:ingestion}

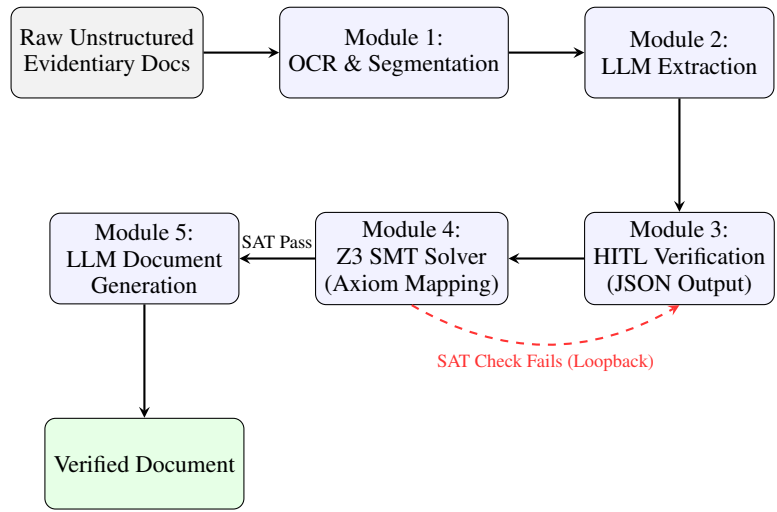
\begin{figure}[htbp]
\centering
\begin{tikzpicture}[
    node distance=1.5cm and 1cm,
    box/.style={rectangle, draw, rounded corners, minimum width=2.5cm, minimum height=1.2cm, align=center, fill=blue!5, font=\small},
    arrow/.style={->, >=stealth, thick},
    loopback/.style={->, >=stealth, thick, dashed, red!80},
    module_label/.style={font=\bfseries\small, text=blue!70!black}
]

\node (raw) [box, fill=gray!10] {Raw Unstructured\\Evidentiary Docs};
\node (mod1) [box, right=of raw] {Module 1:\\OCR \& Segmentation};
\node (mod2) [box, right=of mod1] {Module 2:\\LLM Extraction};
\node (mod3) [box, below=of mod2] {Module 3:\\HITL Verification\\(JSON Output)};
\node (mod4) [box, left=of mod3] {Module 4:\\Z3 SMT Solver\\(Axiom Mapping)};
\node (mod5) [box, left=of mod4] {Module 5:\\LLM Document\\Generation};
\node (out) [box, fill=green!10, below=of mod5] {Verified Document};

\draw [arrow] (raw) -- (mod1);
\draw [arrow] (mod1) -- (mod2);
\draw [arrow] (mod2) -- (mod3);
\draw [arrow] (mod3) -- (mod4);

\draw [loopback] (mod4.south) to[bend right=30] node[midway, below, font=\scriptsize, text=red!80] {SAT Check Fails (Loopback)} (mod3.south);

\draw [arrow] (mod4) -- node[above, font=\scriptsize] {SAT Pass} (mod5);
\draw [arrow] (mod5) -- (out);

\end{tikzpicture}
\caption{End-to-End Neuro-Symbolic Pipeline.}
\label{fig:pipeline}
\end{figure}

The primary stage of the proposed neuro-symbolic pipeline focuses on the  ingestion and structural normalization of unstructured legal text. 
Criminal cases involving summary proceedings for traffic offenses require parsing diverse document types generated across municipal police departments and prosecution offices. 
Module 1 handles four primary evidentiary documents:

\begin{enumerate}
    \item \textbf{DUI Circumstantial Statement Reports} (\textit{juchiwunjeonja jeonghwangjinsulbogoseo} / 주취운전자 정황진술보고서): Containing qualitative observation data regarding a suspect's motor skills, speech pattern, skin flushing, and initial driving behavior.
    \item \textbf{Breathalyzer Test Results} (\textit{eumjucheukjeonggyeolgaji} / 음주측정결과지): Offering the quantitative baseline for legal categorization via blood alcohol concentration (BAC) measurements.
    \item \textbf{Suspect Interrogation Records} (\textit{piuija sinmunjoseo} / 피의자 신문조서): Comprising multi-turn natural language dialogues detailing statements, admissions of guilt, or claims of mitigating circumstances.
    \item \textbf{Traffic Accident Reports} (\textit{gyotongsago balsaengbogoseo} / 교통사고 발생보고서): Archiving spatial-temporal contexts, property damage details, and preliminary evaluations of bodily injuries.
\end{enumerate}

To transform these heterogenous, layout-heavy documents into clean text streams optimized for subsequent Large Language Model (LLM) processing, we implement a hybrid layout-aware ingestion protocol. First, custom rule-based Optical Character Recognition (OCR) is deployed to extract structural text blocks from scanned PDF or image-based inputs. Because legal records are frequently embedded with tabular indices and multi-column administrative metadata, we utilize a bidirectional encoder text segmentation approach to isolate relevant narrative sections while discarding layout noise.

Following initial text extraction, regex-guided field conversion and rule-based table structuring can be applied to systematically format multi-column components—such as victim injury counts or specific vehicle types—into unified, linear strings. This structural normalization neutralizes structural layout fragmentation, ensuring that the downstream semantic parsing components receive coherent, sequential text streams.

%% file: sections/04_extraction.tex

\section{LLM Extraction and Human-in-the-Loop Verification (Modules 2 \& 3)}
\label{sec:extraction}

This section details the continuous workflow bridging the neural extraction of case facts with the mandatory human validation required for legally binding judgments.

\subsection{LLM-based Sentencing Factor Extraction}

Following the ingestion and segmentation of raw evidentiary documents in Module 1, the pipeline employs an LLM as a semantic parser to identify core case elements. The model analyzes the unstructured text to extract primary crime types, continuous quantitative variables, and discrete categorical variables, specifically the special and general sentencing factors defined by the 2026 Sentencing Guidelines for Crimes Related to Traffic Offenses \cite{sc_korea_traffic_2026}.

To mitigate generative hallucination, this extraction process enforces legal syllogism and grounding. The LLM is structurally constrained to map every extracted sentencing factor directly to its exact textual source, designated as the `Reference Text'. 

The extracted entities are systematically populated into an intermediate JSON format. This structured schema acts as the foundational bridge between the probabilistic neural extraction phase and the deterministic formal logic phase. The example provided in Listing \ref{lst:schema} focuses specifically on a Simple DUI (\textit{dansun eumjuunjeon} / 단순 음주운전) offense, demonstrating how quantitative thresholds (e.g., a BAC of 0.12) and qualitative aggravating factors (e.g., high risk to public traffic safety) are isolated for verification:

\begin{lstlisting}[language=json, caption={Example of intermediate JSON schema focusing on a Simple DUI case, containing LLM-extracted factors and flags for human validation.}, label={lst:schema}]
{
  "case_metadata": {
    "crime_types": ["DUI"],
    "blood_alcohol_concentration": 0.12
  },
  "sentencing_factors": {
    "special_factors": {
      "act_aggravating": [
        {
          "factor_name": "High risk to public traffic safety",
          "extracted": true,
          "reference_text": "Incident Report Page 2: Crashed into median barrier",
          "is_valid_by_human": false
        }
      ]
    }
  }
}
\end{lstlisting}

\subsection{Human-in-the-Loop (HITL) Verification}

\begin{figure}[htbp]
\centering
\begin{tikzpicture}[
    box/.style={rectangle, draw, minimum width=6cm, minimum height=7cm, align=left, anchor=north},
    highlight/.style={fill=yellow!40, text=black},
    jsonbox/.style={font=\ttfamily\scriptsize, align=left}
]

\node (leftpanel) [box, fill=gray!5] at (-3.2, 0) {};
\node (rightpanel) [box, fill=blue!5] at (3.2, 0) {};

\node[anchor=north west, font=\bfseries] at (leftpanel.north west) [yshift=-0.2cm, xshift=0.2cm] {Raw Evidentiary Document};
\node[anchor=north west, text width=5.5cm, font=\small] at (leftpanel.north west) [yshift=-1cm, xshift=0.2cm] {
    \textbf{Incident Report (Page 2)}\\[0.3cm]
    At 23:45, the suspect was apprehended.\\[0.2cm]
    \tikz{\node[highlight] {The suspect crashed into the median barrier};} 
    causing significant damage.\\[0.2cm]
    Breathalyzer confirmed \tikz{\node[highlight] {BAC of 0.12\%};}.
};

\node[anchor=north west, font=\bfseries] at (rightpanel.north west) [yshift=-0.2cm, xshift=0.2cm] {Module 3: HITL Verification};
\node[anchor=north west, text width=5.5cm, jsonbox] at (rightpanel.north west) [yshift=-1cm, xshift=0.2cm] {
    \{\\
    \hspace{0.2cm}"blood\_alcohol\_concentration": 0.12,\\
    \hspace{0.2cm}"is\_valid\_by\_human": \textbf{[TRUE]} $\leftarrow$ \textit{User Toggle}\\[0.3cm]
    
    \hspace{0.2cm}"special\_factors": \{\\
    \hspace{0.4cm}"act\_aggravating": [\\
    \hspace{0.6cm}\{\\
    \hspace{0.8cm}"factor\_name": "High risk...",\\
    \hspace{0.8cm}"reference\_text": "Crashed into...",\\
    \hspace{0.8cm}"is\_valid\_by\_human": \textbf{[FALSE]} $\leftarrow$ \textit{Pending}\\
    \hspace{0.6cm}\}\\
    \hspace{0.4cm}]\\
    \hspace{0.2cm}\}\\
    \}
};

\draw[thick, dashed, draw=red!70, -{Stealth}] (-1.5, -2) to[out=0, in=180] (1.5, -4.2);

\end{tikzpicture}
\caption{Wireframe of the HITL interface anchoring extracted variables to raw reference texts.}
\label{fig:hitl_ui}
\end{figure}
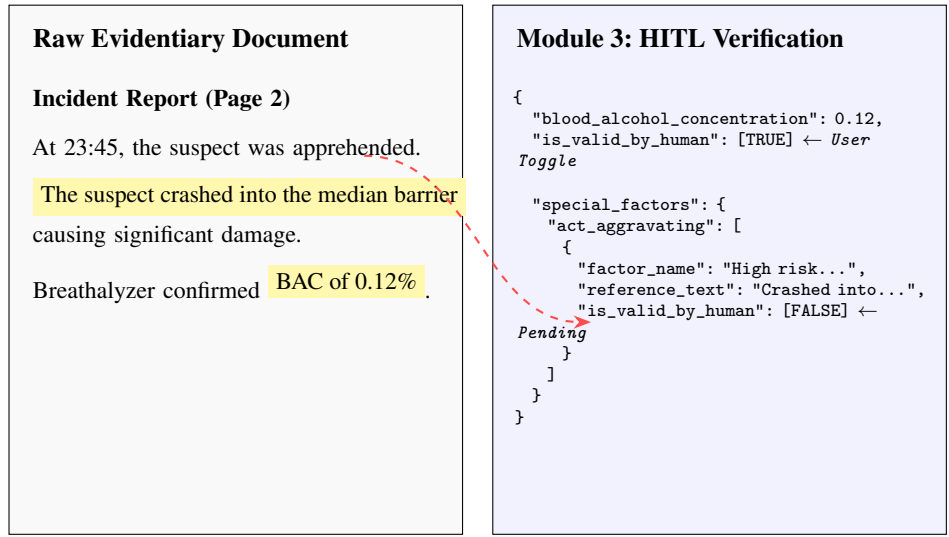

In safety-critical legal environments, relying exclusively on automated extraction introduces risks regarding due process and accountability. To guarantee legal compliance, Module 3 introduces a critical Human-in-the-Loop (HITL) verification phase before any formal constraint calculation occurs.

The verification interface can utilize a split-screen UI design tailored for prosecutorial review. 
On one side, the system presents the structured JSON outputs—specifically the extracted sentencing factors for the DUI charge. 
On the opposing side, it renders the original ingested documents with the corresponding `Reference Text' visually highlighted. 
This layout empowers the legal practitioner to rapidly cross-examine the model's semantic grounding against the raw evidentiary record without manually scanning pages of text.

Crucially, all extracted variables are initialized by default with a boolean flag set to \texttt{is\_valid\_by\_human = false}. 
A legal practitioner must explicitly verify each extracted factor to toggle this state to \texttt{true}. 
Only parameters validated through human oversight are compiled and forwarded to the Z3 SMT solver for the final sentencing computation. 
This structural constraint directly addresses transparency concerns inherent to large language models, positioning the LLM as a verifiable decision-support tool while ensuring that final legal authority remains with the human prosecutor.

%% file: sections/05_solver.tex

\section{Logic Translation and Z3 SMT Solver Execution (Module 4)}
\label{sec:solver}

The core neuro-symbolic bridge of our architecture resides in Module 4, which translates the semantic entities extracted and validated in the previous stages into formal mathematical proofs. Crucially, the system enforces a strict data provenance constraint: only the parameters explicitly authenticated by the legal practitioner (\texttt{is\_valid\_by\_human = true}) in Module 3 are mapped into Boolean and Integer constraints for the Z3 SMT Solver. This guarantees that the formal verification engine computes penalties based on legally accountable facts.

\subsection{Variable Declarations and Axiom Modeling}

Within the Z3 environment, the human-approved JSON fields are initialized as formal state variables. 
For a Simple DUI (\textit{dansun eumjuunjeon} / 단순 음주운전) offense under the Road Traffic Act, the core variables include continuous metrics such as $BAC \in \mathbb{R}$ (Blood Alcohol Concentration) and integer constraints like $Prior\_DUI\_5Y\_Count \in \mathbb{Z}$ (number of similar prior convictions within 5 years). Mitigating and aggravating factors—such as driving with a very low risk to traffic safety (\texttt{SpMit\_LowRisk}) or causing high risk (\texttt{SpAgg\_HighRisk})—are instantiated as Boolean flags.

The 2026 Sentencing Guidelines are subsequently modeled as a set of mathematical axioms. For instance, the automatic statutory categorization based on the BAC threshold is formulated as a conditional constraint. 
An example of this formal mathematical constraint modeled in Z3 for Type 3 BAC categorization is formulated as follows:

$$0.08 \le \text{BAC} < 0.20 \implies \text{Crime\_Category} = 3$$

\begin{figure}[htbp]
\centering
\begin{tikzpicture}[
    node distance=2cm,
    data/.style={rectangle, draw, fill=blue!10, rounded corners, minimum width=3cm, minimum height=1cm, align=center, font=\small},
    process/.style={diamond, draw, fill=orange!20, aspect=2, align=center, font=\small},
    logic/.style={rectangle, draw, fill=purple!10, minimum width=4cm, minimum height=1cm, align=center, font=\ttfamily\small},
    arrow/.style={->, >=stealth, thick},
    loopback/.style={->, >=stealth, thick, dashed, red!80}
]

\node (input) [data] {Verified JSON Payload\\(Continuous \& Boolean Variables)};
\node (translation) [process, below=of input] {SMT Mapping};
\node (z3vars) [logic, below=of translation] {BAC $\in \mathbb{R}$, SpAgg\_HighRisk $\in \mathbb{B}$};
\node (axioms) [process, below=of z3vars] {Z3 SAT Check};
\node (output) [data, fill=green!15, below=of axioms] {Deterministic Fine Range};

\draw [arrow] (input) -- node[right, font=\footnotesize, align=left] {Programmatically\\Parsed} (translation);
\draw [arrow] (translation) -- node[right, font=\footnotesize, align=left] {Cast to Z3\\Variables} (z3vars);
\draw [arrow] (z3vars) -- (axioms);
\draw [arrow] (axioms) -- node[right, font=\footnotesize] {SAT Pass (Math Proof)} (output);

\node (rules) [logic, right=1.5cm of axioms, text width=3.5cm, fill=gray!10] {IF SpAgg > SpMit\\THEN Zone = Aggravated};
\draw [arrow, dashed] (rules) -- (axioms);

\draw [loopback] (axioms.west) -- ++(-1.5,0) |- node[near start, left, font=\footnotesize, text=red!80, align=center] {UNSAT Failure\\(Loopback to HITL)} (input.west);

\end{tikzpicture}
\caption{Data provenance flow mapping deterministic, human-verified JSON outputs to formal SMT logic constraints, including a structural fallback for logical contradictions.}
\label{fig:provenance_flow}
\end{figure}
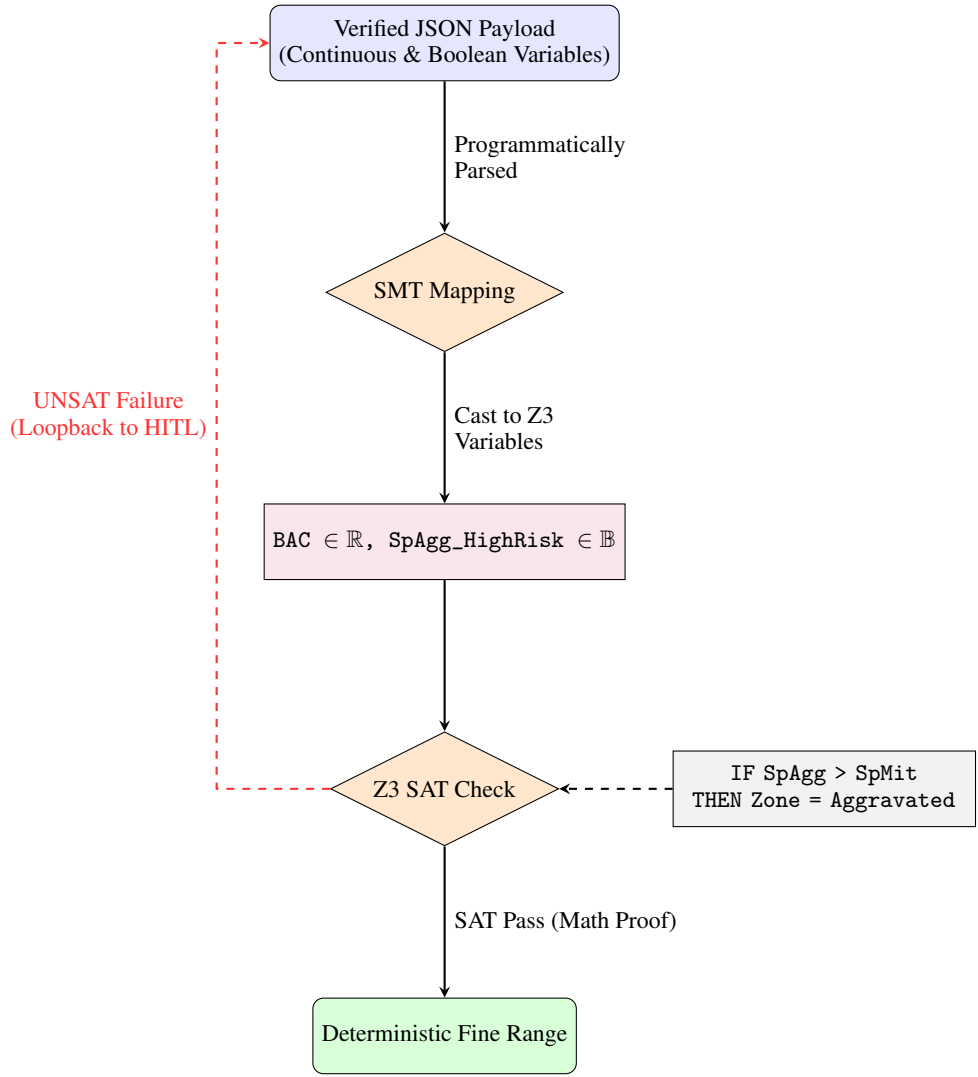

\subsection{Zone Derivation and Mandatory Imprisonment Logic}

Once the base category is established, Z3 evaluates the Boolean arrays of special mitigating and aggravating factors. These Booleans are cast to integers and summed ($Count\_SpMit$ and $Count\_SpAgg$). The solver determines the sentencing zone (Basic, Mitigated, or Aggravated) based on strict inequalities. For example, if $Count\_SpAgg > Count\_SpMit$, the constraint dictates an "Aggravated" zone.

Simultaneously, the solver evaluates mandatory imprisonment constraints, which override monetary fine calculations. For a Type 3 DUI, the logic dictates that imprisonment is mandatory if $Prior\_DUI\_5Y\_Count \ge 3$ or if the zone is ``Aggravated'' and the aggravating factors are exceptionally dominant. These conditions are formulated as composite logical \texttt{OR} and \texttt{AND} constraints, ensuring that repeat offenders are categorically blocked from summary fine proceedings. 

\begin{algorithm}[tbp]
\caption{SMT Constraint Generation Logic for Simple DUI (Road Traffic Act) - 2026 Guidelines}
\label{alg:dui_smt_logic}
\begin{algorithmic}[1]
\REQUIRE $BAC \in \mathbb{R}$, $Prior\_DUI\_5Y\_Count \in \mathbb{Z}$, $Has\_Any\_Prior\_DUI \in \mathbb{B}$
\REQUIRE $SpMit\_Act \in \mathbb{B}^a, SpMit\_Actor \in \mathbb{B}^b$ \COMMENT{Bifurcated mitigating factors}
\REQUIRE $SpAgg\_Act \in \mathbb{B}^x, SpAgg\_Actor \in \mathbb{B}^y$ \COMMENT{Bifurcated aggravating factors}
\REQUIRE $NP\_Exception \in \mathbb{B}$ \COMMENT{Non-Punishment / Damage Recovery flag}

\STATE \textbf{// 1. Target Type Mapping based on BAC}
\IF{$0.03 \le BAC < 0.08$}
    \STATE $DUI\_Type \leftarrow 2$
\ELSIF{$0.08 \le BAC < 0.20$}
    \STATE $DUI\_Type \leftarrow 3$
\ELSIF{$BAC \ge 0.20$}
    \STATE $DUI\_Type \leftarrow 4$
\ELSE
    \STATE $DUI\_Type \leftarrow 0$
\ENDIF

\STATE \textbf{// 2. Quantify Special Factors and Apply Statutory Exception}
\STATE \COMMENT{NP\_Exception is elevated from Actor to Act tier for calculation}
\STATE $Count\_Mit\_Act \leftarrow \sum \text{Int}(SpMit\_Act) + \text{Int}(NP\_Exception)$
\STATE $Count\_Mit\_Actor \leftarrow \sum \text{Int}(SpMit\_Actor) - \text{Int}(NP\_Exception)$
\STATE $Count\_Agg\_Act \leftarrow \sum \text{Int}(SpAgg\_Act)$
\STATE $Count\_Agg\_Actor \leftarrow \sum \text{Int}(SpAgg\_Actor)$

\STATE $Total\_Mit \leftarrow Count\_Mit\_Act + Count\_Mit\_Actor$
\STATE $Total\_Agg \leftarrow Count\_Agg\_Act + Count\_Agg\_Actor$

\STATE \textbf{// 3. Zone Derivation (Act Factor Superiority Principle Applied)}
\IF{$Total\_Agg > Total\_Mit$}
    \STATE $Zone \leftarrow \text{"Aggravated"}$
\ELSIF{$Total\_Mit > Total\_Agg$}
    \STATE $Zone \leftarrow \text{"Mitigated"}$
\ELSE
    \STATE \COMMENT{Tie-breaker: Act factors hold superior weight}
    \IF{$Count\_Agg\_Act > Count\_Mit\_Act$}
        \STATE $Zone \leftarrow \text{"Aggravated"}$
    \ELSIF{$Count\_Mit\_Act > Count\_Agg\_Act$}
        \STATE $Zone \leftarrow \text{"Mitigated"}$
    \ELSE
        \STATE $Zone \leftarrow \text{"Basic"}$
    \ENDIF
\ENDIF

\STATE \textbf{// 4. Special Dominance Conditions (Based on Totals)}
\STATE $Is\_SpAgg\_Dominant \leftarrow (Total\_Mit == 0 \land Total\_Agg \ge 2) \lor (Total\_Agg - Total\_Mit \ge 2)$
\STATE $Is\_SpMit\_Dominant \leftarrow (Total\_Agg == 0 \land Total\_Mit \ge 2) \lor (Total\_Mit - Total\_Agg \ge 2)$

\STATE \textbf{// 5. Mandatory Imprisonment Check (Fine Calculation Override)}
\STATE $Prison\_Type2 \leftarrow DUI\_Type == 2 \land (Prior\_DUI\_5Y\_Count \ge 3 \lor (Zone == \text{"Aggravated"} \land Is\_SpAgg\_Dominant))$
\STATE $Prison\_Type3 \leftarrow DUI\_Type == 3 \land (Prior\_DUI\_5Y\_Count \ge 3 \lor (Zone == \text{"Aggravated"} \land Is\_SpAgg\_Dominant))$
\STATE $Prison\_Type4 \leftarrow DUI\_Type == 4 \land (Prior\_DUI\_5Y\_Count \ge 3 \lor (Zone == \text{"Aggravated"} \land (Has\_Any\_Prior\_DUI \lor Is\_SpAgg\_Dominant)))$
\STATE $Recommend\_Prison \leftarrow Prison\_Type2 \lor Prison\_Type3 \lor Prison\_Type4$

\STATE \textbf{// 6. Base Fine Range Calculation \& 7. Final Adjustment (KRW 10,000s)}
\IF{$Recommend\_Prison == \text{True}$}
    \RETURN $\text{"Imprisonment\_Recommended"}$
\ELSE
    \STATE $Base\_Range \leftarrow \text{LookupFine}(DUI\_Type, Zone)$
    \STATE $Final\_Min \leftarrow Is\_SpMit\_Dominant \text{ ? } Base\_Range.Min \times 0.5 \text{ : } Base\_Range.Min$
    \STATE $Final\_Max \leftarrow Is\_SpAgg\_Dominant \text{ ? } Base\_Range.Max \times 1.5 \text{ : } Base\_Range.Max$
    \RETURN $[Final\_Min, Final\_Max]$
\ENDIF
\end{algorithmic}
\end{algorithm}

\subsection{Satisfiability Check and Deterministic Output}

Before generating a final value, Z3 executes a Satisfiability (SAT) check across the entire axiom set. This step mathematically proves that no logical contradictions exist within the applied legal rules (e.g., ensuring a single case cannot simultaneously evaluate to both "Mitigated" and "Aggravated" fine ranges). 

If the solver encounters an unsatisfiable state (\texttt{UNSAT})—for instance, if mutually exclusive sentencing variables were accidentally validated by the human user during Module 3—the execution halts. To ensure robust exception handling, the pipeline triggers a structural fallback mechanism that loops back to the HITL interface. This error payload highlights the contradictory variables, requiring the legal practitioner to resolve the logical conflict before the formal calculation can proceed.

If the SAT check passes and imprisonment is not mandatory, the solver computes the deterministic `Recommended Fine Range (Min-Max)'. It applies zone-specific baselines (e.g., $5,000,000$ to $8,000,000$ KRW for a Basic Type 3 DUI) and executes final adjustments, such as halving the minimum fine if mitigating factors are overwhelmingly dominant, or multiplying the maximum by 1.5 if aggravating factors dominate. 
Because this output is derived from an automated theorem prover rather than a LLM, the resulting fine range is guaranteed to be free of hallucination.

%% file: sections/06_generation.tex

\section{Visualization and Document Generation (Module 5)}
\label{sec:generation}

The final module of our neuro-symbolic framework focuses on translating the deterministic mathematical outputs of the Z3 SMT solver into actionable artifacts for legal practitioners. 
This ensures that the system serves as an accountable assistive tool rather than a decision-maker.

\subsection{Sentencing Tree Visualization}

While the Z3 SMT solver guarantees mathematical correctness and logical satisfiability, raw Boolean execution traces and algebraic constraints are practically inscrutable to the average prosecutor or judge. To achieve true explanation-centric auditability, the pipeline automatically renders the Z3 deterministic sentencing calculation into a visual logic tree.

This visualization depicts the exact hierarchical decision nodes, beginning with the initial Blood Alcohol Concentration (BAC) categorization, traversing the quantified mitigating and aggravating factors, and concluding in either a final fine calculation or a mandatory imprisonment flag. 
By explicitly highlighting the active legal pathway---such as illustrating how a Type 3 DUI is mitigated by verified special circumstances---the system provides a transparent, step-by-step rationale for the Z3 solver's outputs. 
Should a prosecutor need to substantiate a summary indictment during an appeal, this visual decision tree serves as an interpretable audit trail that directly connects statutory logic with verified case evidence.

\begin{figure}[htbp]
    \centering
    \begin{tikzpicture}[
        node distance=2cm,
        auto,
        box/.style={rectangle, draw, rounded corners, fill=blue!5, text width=4.5cm, text centered, minimum height=1cm, font=\small},
        decision/.style={diamond, aspect=2, draw, fill=orange!5, text width=3cm, text centered, minimum height=1cm, font=\small},
        result/.style={rectangle, draw=green!60!black, very thick, rounded corners, fill=green!5, text width=5.5cm, text centered, minimum height=1.2cm, font=\bfseries\small},
        line/.style={draw, thick, -latex}
    ]
    
    \node [box] (start) {Start: Verified DUI Case Facts};
    \node [decision, below=0.6cm of start] (bac) {BAC Level Check};
    \node [box, below=0.8cm of bac] (type3) {Type 3 DUI};
    \node [decision, below=0.6cm of type3] (spec) {Evaluate Factors};
    \node [box, below=1.2cm of spec] (zone) {Mitigated Zone};
    \node [decision, below=0.6cm of zone] (prison) {Mandatory Imprisonment Check};
    \node [box, below=1.4cm of prison] (base) {Base Fine Range:\\ 2,000,000 - 4,000,000 KRW};
    \node [decision, below=0.6cm of base] (dom) {Dominance Adjustment};
    \node [result, below=1.2cm of dom] (final) {Final Recommended Fine:\\ 1,000,000 - 4,000,000 KRW};

    \path [line] (start) -- (bac);
    \path [line] (bac) -- node[right, font=\footnotesize] {BAC = 0.12} (type3);
    \path [line] (type3) -- (spec);
    \path [line] (spec) -- node[right, font=\footnotesize, text width=4.5cm] {Mitigating: Low Risk (1)\\Aggravating: None (0)\\(Count\_SpMit > Count\_SpAgg)} (zone);
    \path [line] (zone) -- (prison);
    \path [line] (prison) -- node[right, font=\footnotesize, text width=4.5cm] {Prior Convictions $< 3$\\No Dominant Aggravation} (base);
    \path [line] (base) -- (dom);
    \path [line] (dom) -- node[right, font=\footnotesize, text width=4.5cm] {\texttt{Is\_SpMit\_Dominant = True}\\(Min Fine $\times$ 0.5)} (final);
    
    \end{tikzpicture}
    \caption{Visualization of the deterministic sentencing tree for a Type 3 Simple DUI with a dominant mitigating factor.}
    \label{fig:sentencing_tree}
\end{figure}
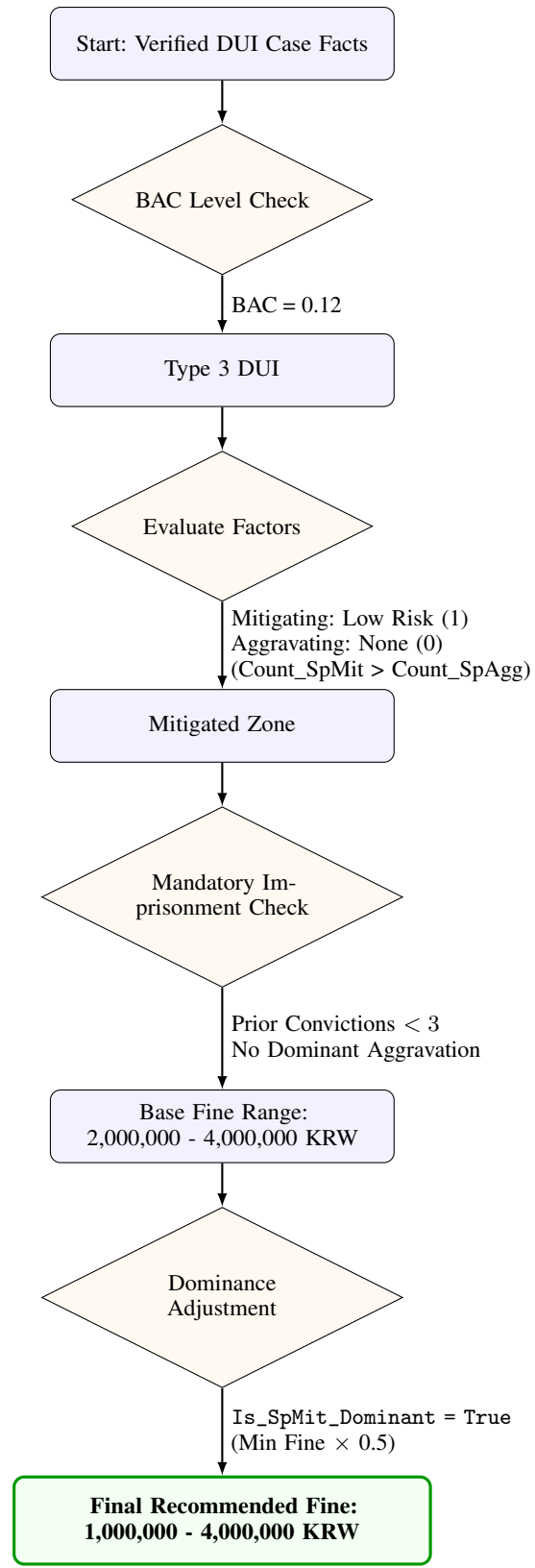

\subsection{Prompt Engineering for Legal Document Generation}
\label{subsec:prompts}

To ensure the LLM generates documents that strictly adhere to South Korean prosecutorial formatting, we utilize rigidly structured prompts. 
These prompts explicitly define the mandatory sections for each specific legal document and inject the mathematically verified outputs from the Z3 SMT solver.

Below are the system prompt templates utilized in Module 5 for drafting the Police Investigation Report (\textit{susa-bogoseo} / 수사보고서), the Police Forwarding Opinion (\textit{songchi-uikyeonseo} / 송치의견서), and the Summary Indictment Request (\textit{yaksik-gongsojang} / 약식공소장).

\begin{tcolorbox}[colback=gray!5, colframe=gray!50, title=\textbf{System Prompt for Police Investigation Report (\textit{susa-gyeolgwabogoseo} / 수사결과보고서)}]
\textbf{System:} You are an expert South Korean police investigator. \\
\textbf{Task:} Draft a ``Police Investigation Report'' (\textit{susa-gyeolgwabogoseo} / 수사결과보고서) based strictly on the provided JSON case facts. \\
\textbf{Constraints:} You must not invent any facts. You must include the following 7 mandatory sections exactly as formatted below:
\begin{enumerate}
    \item Suspect Information (\textit{piuija injeoksahang} / 피의자 인적사항)
    \item Criminal Record (\textit{beomjoegyeongryeokjaryo} / 범죄경력자료)
    \item Facts of the Crime (\textit{beomjoesasiri} / 범죄사실 - Use the 5W1H principle)
    \item Applicable Statutory Provisions (\textit{jeogyongbeopjo} / 적용법조)
    \item Evidence (\textit{jeungeogwangye} / 증거관계 - Split into Personal and Material Evidence)
    \item Investigation Result and Opinion (\textit{susagyeolgwa mit uigyeon} / 수사결과 및 의견)
    \item Participating Officer (\textit{susa-chamyeogyeongchalgoan} / 수사참여경찰관)
\end{enumerate}
\textbf{Input Data:} \{Verified\_JSON\_Payload\}
\end{tcolorbox}

\begin{tcolorbox}[colback=gray!5, colframe=gray!50, title=\textbf{System Prompt for Police Forwarding Opinion (\textit{songchi-uigyeonseo} / 송치의견서)}]
\textbf{System:} You are an expert South Korean police investigator preparing a case for prosecutorial review. \\
\textbf{Task:} Draft a ``Police Forwarding Opinion'' (\textit{songchi-uigyeonseo} / 송치의견서) recommending Indictment (\textit{giso-uigyeon} / 기소의견). \\
\textbf{Constraints:} The tone must be objective and legally formal. Maintain absolute factual consistency with the Investigation Report. Structure the output into the following 5 mandatory sections:
\begin{enumerate}
    \item Suspect Information (\textit{piuija injeoksahang} / 피의자 인적사항 - Mask RRN as [RRN Omitted])
    \item Charge and Applicable Law (\textit{joemyeong mit jeogyongbeopjo} / 죄명 및 적용법조)
    \item Facts of the Crime (\textit{beomjoesasiri} / 범죄사실)
    \item Evidence List (\textit{jeungeogwangye} / 증거관계)
    \item Measures and Forwarding Opinion (\textit{jochi mit susa-uigyeon} / 조치 및 수사의견 - Must conclude with ``Indictment Opinion'' (\textit{giso-uigyeon} / 기소의견) based on the evidence.)
\end{enumerate}
\textbf{Input Data:} \{Verified\_JSON\_Payload\}
\end{tcolorbox}

\begin{tcolorbox}[colback=gray!5, colframe=gray!50, title=\textbf{System Prompt for Summary Indictment Request (\textit{yaksik-gongsojang} / 약식공소장)}]
\textbf{System:} You are an expert South Korean Prosecutor drafting a summary indictment. \\
\textbf{Task:} Draft a ``Summary Indictment Request'' (\textit{yaksik-gongsojang} / 약식공소장). \\
\textbf{CRITICAL CONSTRAINT:} You must extract the exact recommended fine from the \texttt{Z3\_Solver\_Output} and place it verbatim into the ``Prosecutor's Demand'' (\textit{guhyeongryang} / 구형량) section. DO NOT calculate or modify this number.

\textbf{Structure the document exactly as follows:}
\begin{itemize}
    \item \textbf{[ Indictment and Summary Order Request ]} (\textit{gongsojang mit yaksikmyeongnyeong-cheongguseo} / 공소장 및 약식명령청구서)
    \item Case, Defendant, Residence, Charge, Applicable Law (\textit{sajeon, pigoin, jugeo, joemyeong, jeogyongbeopjo} / 사건, 피고인, 주거, 죄명, 적용법조)
    \item \textbf{[ Facts of the Crime ]} (\textit{gongsosasiri} / 공소사실)
    \begin{itemize}
        \item Draft the chronological narrative of the crime based on the JSON.
    \end{itemize}
    \item \textbf{[ Prosecutor's Demand and Intent of Claim ]} (\textit{guhyeongryang mit cheonggu-ui chwiji} / 구형량 및 청구의 취지)
    \begin{itemize}
        \item ``The defendant is subject to a fine of \{\texttt{Z3\_Solver\_Output.Final\_Fine}\} KRW.'' (피고인을 벌금 \{\texttt{Z3\_Solver\_Output.Final\_Fine}\}원에 처한다.)
        \item ``Accordingly, a summary order is requested pursuant to Article 448 of the Criminal Procedure Act.'' (이에 형사소송법 제448조에 의하여 약식명령을 청구합니다.)
    \end{itemize}
\end{itemize}
\textbf{Input Data:} \{Verified\_JSON\_Payload\} \\
\textbf{Solver Data:} \{Z3\_Solver\_Output\}
\end{tcolorbox}

%% file: sections/07_conclusion.tex

\section{Conclusion}
\label{sec:conclusion}

\subsection{Summary of Contributions}
In this paper, we introduced a neuro-symbolic framework designed to automate the drafting of summary indictments (\textit{guyaksik} / 구약식) in the Korean prosecutorial system. 
While LLMs excel at semantic extraction from unstructured legal texts, 
their probabilistic nature introduces risks of hallucination in high-stakes statutory calculations. 

To resolve this limitation, our architecture strictly confines the LLM to a data extraction role and delegates the complex arithmetic of the 2026 Sentencing Guidelines to a formal Z3 Satisfiability Modulo Theories (SMT) solver. 
By enforcing an intermediate Human-in-the-Loop (HITL) verification schema, the system ensures that all deterministic fine calculations are anchored to legally authenticated facts. 
Finally, the framework closes the automation loop by dynamically generating formally compliant legal documents, such as the Summary Indictment Request (\textit{yaksik-gongsojang} / 약식공소장), serving as an accountable, verifiable assistive tool for legal practitioners.

\subsection{Future Work}
Because the current study focuses on the theoretical formulation and logical mapping of this neuro-symbolic architecture, the immediate next phase of research will center on full-scale software implementation. 
To realize this goal, existing specialized components can be integrated directly into our pipeline. 
Specifically, the data-efficient generative extraction techniques developed by Hwang et al. \cite{hwang2022data} can serve as a core engine for Module 2, enabling reliable extraction of structured case facts from unstructured legal records with minimal annotated training data. 
Additionally, the LBox Open benchmark suite and pre-trained resources from Hwang et al. \cite{hwang2022multi} offer a valuable foundation for domain adaptation and system-wide evaluation across Korean legal text processing tasks.
Future empirical studies will evaluate the operationalized system using real-world prosecutorial datasets to measure LLM extraction accuracy, Z3 solver soundness, and the practical time-on-task efficiency gained by prosecutors during manual review.

Furthermore, we plan to expand the framework's mathematical axioms beyond traffic offenses to encompass other minor crimes that predominantly conclude with summary proceedings and monetary fines. These domains include:

\begin{itemize}
    \item \textbf{Violent Crimes} (\textit{pokryeok-beomjoe} / 폭력범죄): Minor infractions such as simple assault (\textit{dansun-pokhaeng} / 단순 폭행) involving minor altercations, and simple injury (\textit{dansun-sanghae} / 단순 상해) with short recovery periods.
    \item \textbf{Defamation Crimes} (\textit{myeongyehoeson-beomjoe} / 명예훼손범죄): Internet defamation and insult (\textit{internet-myeongyehoeson mit moyok} / 인터넷 명예훼손 및 모욕) occurring across online communities and social media platforms.
    \item \textbf{Minor Property Crimes} (\textit{jaesan-beomjoe-ui gyeongmihan-yuyeong} / 재산범죄의 경미한 유형): Incidents of minor theft (\textit{dansun-jeoldo} / 단순 절도) or accidental property damage (\textit{dansun-jaemulsongeoe} / 단순 재물손괴).
    \item \textbf{Electronic Financial Transactions Act Violations} (\textit{jeonjageumyunggeoraebeop-wiban-beomjoe} / 전자금융거래법위반범죄): First-time offenses involving the simple lending or transfer of passbooks and debit cards (\textit{dansun-daeyeo mit yangdo} / 단순 대여 및 양도).
    \item \textbf{Labor Standards Act Violations} (\textit{geunrogijunbeop-wiban-beomjoe} / 근로기준법위반범죄): Unpaid wage cases (\textit{soaek-imgeum deung mijigeup} / 소액 임금 등 미지급) where the employer demonstrates intent to pay or faces extenuating financial hardships.
    \item \textbf{Trespassing Crimes} (\textit{jugeochimbip-beomjoe} / 주거침입범죄): Simple trespassing (\textit{dansun-jugeochimbip} / 단순 주거침입) into common areas or building corridors without further criminal intent.
\end{itemize}

By scaling the deterministic logic engine to accommodate these diverse legal scenarios, this framework aims to alleviate the administrative bottleneck of minor prosecutorial caseloads while preserving judicial integrity.

%% file: sections/08_appendix.tex
\newpage
\appendix
\section{Algorithm for Driving Without a License}
\label{sec:appendix_a}

\begin{algorithm}[H]
\caption{SMT Constraint Generation Logic for Driving Without a License - 2026 Guidelines}
\label{alg:unlicensed_smt_logic}
\begin{algorithmic}[1]
\REQUIRE $Prior\_5Y\_Count \in \mathbb{Z}$ \COMMENT{Similar prior convictions within 5 years}
\REQUIRE $SpMit\_Act \in \mathbb{B}^a, SpMit\_Actor \in \mathbb{B}^b$ \COMMENT{Bifurcated mitigating factors}
\REQUIRE $SpAgg\_Act \in \mathbb{B}^x, SpAgg\_Actor \in \mathbb{B}^y$ \COMMENT{Bifurcated aggravating factors}
\REQUIRE $NP\_Exception \in \mathbb{B}$ \COMMENT{Non-Punishment / Damage Recovery flag}
\REQUIRE $GenMit \in \mathbb{B}^3$ \COMMENT{General factors (e.g., Livelihood); ignored in Zone calculation}

\STATE \textbf{// 1. Quantify Special Factors and Apply Statutory Exception}
\STATE \COMMENT{NP\_Exception is elevated from Actor to Act tier for calculation}
\STATE $Count\_Mit\_Act \leftarrow \sum \text{Int}(SpMit\_Act) + \text{Int}(NP\_Exception)$
\STATE $Count\_Mit\_Actor \leftarrow \sum \text{Int}(SpMit\_Actor) - \text{Int}(NP\_Exception)$
\STATE $Count\_Agg\_Act \leftarrow \sum \text{Int}(SpAgg\_Act)$
\STATE $Count\_Agg\_Actor \leftarrow \sum \text{Int}(SpAgg\_Actor)$

\STATE $Total\_Mit \leftarrow Count\_Mit\_Act + Count\_Mit\_Actor$
\STATE $Total\_Agg \leftarrow Count\_Agg\_Act + Count\_Agg\_Actor$

\STATE \textbf{// 2. Zone Derivation (Act Factor Superiority Principle Applied)}
\IF{$Total\_Agg > Total\_Mit$}
    \STATE $Zone \leftarrow \text{"Aggravated"}$
\ELSIF{$Total\_Mit > Total\_Agg$}
    \STATE $Zone \leftarrow \text{"Mitigated"}$
\ELSE
    \STATE \COMMENT{Tie-breaker: Act factors hold superior weight}
    \IF{$Count\_Agg\_Act > Count\_Mit\_Act$}
        \STATE $Zone \leftarrow \text{"Aggravated"}$
    \ELSIF{$Count\_Mit\_Act > Count\_Agg\_Act$}
        \STATE $Zone \leftarrow \text{"Mitigated"}$
    \ELSE
        \STATE $Zone \leftarrow \text{"Basic"}$
    \ENDIF
\ENDIF

\STATE \textbf{// 3. Special Dominance Conditions (Based on Totals)}
\STATE $Is\_SpAgg\_Dominant \leftarrow (Total\_Mit == 0 \land Total\_Agg \ge 2) \lor (Total\_Agg - Total\_Mit \ge 2)$
\STATE $Is\_SpMit\_Dominant \leftarrow (Total\_Agg == 0 \land Total\_Mit \ge 2) \lor (Total\_Mit - Total\_Agg \ge 2)$

\STATE \textbf{// 4. Mandatory Imprisonment Check (Fine Calculation Override)}
\STATE \COMMENT{Note: For the Mitigated zone, imprisonment for $\ge 3$ priors is optional, not mandatory.}
\STATE $Prison\_Basic \leftarrow Zone == \text{"Basic"} \land Prior\_5Y\_Count \ge 3$
\STATE $Prison\_Aggravated \leftarrow Zone == \text{"Aggravated"} \land (Prior\_5Y\_Count \ge 3 \lor Is\_SpAgg\_Dominant)$
\STATE $Recommend\_Prison \leftarrow Prison\_Basic \lor Prison\_Aggravated$

\STATE \textbf{// 5. Base Fine Range Calculation (KRW 10,000s)}
\IF{$Recommend\_Prison == \text{True}$}
    \RETURN $\text{"Imprisonment\_Recommended"}$
\ELSE
    \IF{$Zone == \text{"Mitigated"}$}
        \STATE $Base\_Range \leftarrow [50, 150]$
    \ELSIF{$Zone == \text{"Basic"}$}
        \STATE $Base\_Range \leftarrow [100, 200]$
    \ELSIF{$Zone == \text{"Aggravated"}$}
        \STATE $Base\_Range \leftarrow [150, 300]$
    \ENDIF
    
    \STATE \textbf{// 6. Final Recommendation Adjustment}
    \STATE $Final\_Min \leftarrow Is\_SpMit\_Dominant \text{ ? } Base\_Range.Min \times 0.5 \text{ : } Base\_Range.Min$
    \STATE $Final\_Max \leftarrow Is\_SpAgg\_Dominant \text{ ? } Base\_Range.Max \times 1.5 \text{ : } Base\_Range.Max$
    
    \RETURN $[Final\_Min, Final\_Max]$
\ENDIF
\end{algorithmic}
\end{algorithm}

Under the Road Traffic Act, Driving Without a License (\textit{mumyeonheo-unjeon} / 무면허운전) occurs when an individual operates a motor vehicle without legally valid driving privileges (e.g., license never obtained, suspended, or revoked). The 2026 Sentencing Guidelines categorize this as a Type 1 offense within the DUI and Unlicensed Driving section. Cases without severe aggravating factors or extensive prior convictions typically result in summary proceedings (\textit{guyaksik} / 구약식) leading to a monetary fine.

To expand the theoretical framework established in Module 4, Algorithm \ref{alg:unlicensed_smt_logic} provides the explicit SMT constraint generation logic for Driving Without a License. It details the instantiation of special mitigating and aggravating factors, zone derivation, and the final deterministic fine calculation. Note that General Mitigating Factors (\textit{ilban-gamgyeong-inja} / 일반감경인자), such as committing the crime for livelihood reasons (\textit{saenggye-hyeong beomjoe} / 생계형 범죄), are recorded for discretionary judicial reference but are strictly excluded from the formal zone derivation math.

\newpage
\section{Algorithm for Traffic Casualties (Bodily Injury)}
\label{sec:appendix_b}

\begin{algorithm}[H]
\caption{SMT Constraint Generation Logic for Traffic Casualties (Bodily Injury) - 2026 Guidelines}
\label{alg:casualties_smt_logic}
\begin{algorithmic}[1]
\REQUIRE $SpMit\_Act \in \mathbb{B}^a, SpMit\_Actor \in \mathbb{B}^b$ \COMMENT{Bifurcated mitigating factors}
\REQUIRE $SpAgg\_Act \in \mathbb{B}^x, SpAgg\_Actor \in \mathbb{B}^y$ \COMMENT{Bifurcated aggravating factors}
\REQUIRE $Minor\_Injury \in \mathbb{B}$ \COMMENT{Extracted specific mitigating factor (formerly SpMit[2])}
\REQUIRE $NP\_Exception \in \mathbb{B}$ \COMMENT{Non-Punishment / Recovery flag (formerly SpMit[6])}
\REQUIRE $GenAgg \in \mathbb{B}^2$ \COMMENT{General factors (e.g., Serious Injury); ignored in Zone calculation}

\STATE \textbf{// 1. Quantify Special Factors and Apply Statutory Exception}
\STATE \COMMENT{NP\_Exception is elevated from Actor to Act tier for calculation}
\STATE $Count\_Mit\_Act \leftarrow \sum \text{Int}(SpMit\_Act) + \text{Int}(NP\_Exception) + \text{Int}(Minor\_Injury)$
\STATE $Count\_Mit\_Actor \leftarrow \sum \text{Int}(SpMit\_Actor) - \text{Int}(NP\_Exception) - \text{Int}(Minor\_Injury)$
\STATE $Count\_Agg\_Act \leftarrow \sum \text{Int}(SpAgg\_Act)$
\STATE $Count\_Agg\_Actor \leftarrow \sum \text{Int}(SpAgg\_Actor)$

\STATE $Total\_Mit \leftarrow Count\_Mit\_Act + Count\_Mit\_Actor$
\STATE $Total\_Agg \leftarrow Count\_Agg\_Act + Count\_Agg\_Actor$

\STATE \textbf{// 2. Zone Derivation (Act Factor Superiority Principle Applied)}
\IF{$Total\_Agg > Total\_Mit$}
    \STATE $Zone \leftarrow \text{"Aggravated"}$
\ELSIF{$Total\_Mit > Total\_Agg$}
    \STATE $Zone \leftarrow \text{"Mitigated"}$
\ELSE
    \STATE \COMMENT{Tie-breaker: Act factors hold superior weight}
    \IF{$Count\_Agg\_Act > Count\_Mit\_Act$}
        \STATE $Zone \leftarrow \text{"Aggravated"}$
    \ELSIF{$Count\_Mit\_Act > Count\_Agg\_Act$}
        \STATE $Zone \leftarrow \text{"Mitigated"}$
    \ELSE
        \STATE $Zone \leftarrow \text{"Basic"}$
    \ENDIF
\ENDIF

\STATE \textbf{// 3. Special Dominance Conditions (Based on Totals)}
\STATE $Is\_SpAgg\_Dominant \leftarrow (Total\_Mit == 0 \land Total\_Agg \ge 2) \lor (Total\_Agg - Total\_Mit \ge 2)$
\STATE $Is\_SpMit\_Dominant \leftarrow (Total\_Agg == 0 \land Total\_Mit \ge 2) \lor (Total\_Mit - Total\_Agg \ge 2)$

\STATE \textbf{// 4. Exceptional Allowance for Fines in the Aggravated Zone}
\STATE $Allow\_Fine\_In\_Aggravated \leftarrow Minor\_Injury \lor NP\_Exception$

\STATE \textbf{// 5. Mandatory Imprisonment Check (Fine Calculation Override)}
\STATE $Prison\_Mandatory\_Aggravated \leftarrow Zone == \text{"Aggravated"} \land (\neg Allow\_Fine\_In\_Aggravated \lor Is\_SpAgg\_Dominant)$

\STATE \textbf{// 6. Base Fine Range Calculation (KRW 10,000s)}
\IF{$Prison\_Mandatory\_Aggravated == \text{True}$}
    \RETURN $\text{"Imprisonment\_Recommended"}$
\ELSE
    \IF{$Zone == \text{"Mitigated"}$}
        \STATE $Base\_Range \leftarrow [100, 700]$
    \ELSIF{$Zone == \text{"Basic"}$}
        \STATE $Base\_Range \leftarrow [500, 1200]$
    \ELSIF{$Zone == \text{"Aggravated"}$}
        \STATE $Base\_Range \leftarrow [800, 2000]$
    \ENDIF
    
    \STATE \textbf{// 7. Final Recommendation Adjustment}
    \STATE $Final\_Min \leftarrow Is\_SpMit\_Dominant \text{ ? } Base\_Range.Min \times 0.5 \text{ : } Base\_Range.Min$
    \STATE $Final\_Max \leftarrow Is\_SpAgg\_Dominant \text{ ? } Base\_Range.Max \times 1.5 \text{ : } Base\_Range.Max$
    
    \RETURN $[Final\_Min, Final\_Max]$
\ENDIF
\end{algorithmic}
\end{algorithm}

Under the Act on Special Cases Concerning the Settlement of Traffic Accidents (\textit{gyotong-sago-cheori-teungnye-beop} / 교통사고처리특례법), Traffic Casualties (\textit{gyotongsago chisang} / 교통사고 치상) occur when a driver causes bodily injury to another person through professional or gross negligence. The 2026 Sentencing Guidelines categorize this as a Type 1 General Traffic Accident. 

While bodily injury cases frequently proceed to formal trials, summary proceedings (\textit{guyaksik} / 구약식) leading to a monetary fine are often permitted if the injuries are demonstrably minor (\textit{gyeongmi-han sanghae} / 경미한 상해) or if the victim explicitly expresses no desire for punishment and substantial damage recovery has occurred (\textit{cheobeol-bulwon ttoneun siljiljeok pihae-hoebok} / 처벌불원 또는 실질적 피해회복). 

Algorithm \ref{alg:casualties_smt_logic} models the SMT constraint logic for this specific offense. Notably, it demonstrates how formal logic handles conditional exceptions: even if the mathematical zone is evaluated as ``Aggravated'' (which defaults to mandatory imprisonment), the solver is programmed to exceptionally allow a monetary fine if specific overriding mitigating factors are validated, provided the aggravating factors are not mathematically dominant.